\documentclass[aps,prl,showpacs,twocolumn,superscriptaddress]{revtex4}
\usepackage{amsmath,bm}
\usepackage{graphicx,color}
\newcommand{\ket}[1]{| #1 \rangle}
\newcommand{\cre}[1]{d_{#1}^{\dagger}}
\newcommand{\ann}[1]{d_{#1}}
\newcommand{\hc}{{\rm h.c.}}
\newcommand{\ep}{\epsilon_p}
\newcommand{\ed}{\epsilon_d}
\newcommand{\pps}{(pp\sigma)}
\newcommand{\ppp}{(pp\pi)}
\newcommand{\pds}{(pd\sigma)}
\newcommand{\pdp}{(pd\pi)}
\begin{document}
\title{Quantum spin Hall effect in a transition metal oxide Na$_2$IrO$_3$}
\author{Atsuo Shitade}
\email{shitade@appi.t.u-tokyo.ac.jp}
\affiliation{Department of Applied Physics, The University of Tokyo, Hongo, Bunkyo-ku, Tokyo 113-8656, Japan}
\author{Hosho Katsura}
\affiliation{Cross-Correlated Materials Research Group (CMRG), ASI, RIKEN, Wako 351-0198, Japan}
\author{Jan Kune{\v s}}
\affiliation{Theoretical Physics III, Center for Electronic Correlations and Magnetism, Institute of Physics, University of Augsburg, Augsburg 86135, Germany}
\affiliation{Institute of Physics, Academy of Sciences of the Czech Republic, Cukrovarnick{\' a} 10, 162 53 Praha, Czech Republic}
\author{Xiao-Liang Qi}
\affiliation{Department of Physics, McCullough Building, Stanford University, Stanford, CA 94305-4045, USA}
\author{Shou-Cheng Zhang}
\affiliation{Department of Physics, McCullough Building, Stanford University, Stanford, CA 94305-4045, USA}
\author{Naoto Nagaosa}
\affiliation{Department of Applied Physics, The University of Tokyo, Hongo, Bunkyo-ku, Tokyo 113-8656, Japan}
\affiliation{Cross-Correlated Materials Research Group (CMRG), ASI, RIKEN, Wako 351-0198, Japan}
\begin{abstract}
We study theoretically the electronic states in a $5d$ transition metal oxide Na$_2$IrO$_3$, 
in which both the spin-orbit interaction and the electron correlation play crucial roles.
Tight-binding model analysis together with the fisrt-principles band structure calculation
predicts that this material is a layered quantum spin Hall system.
Due to the electron correlation, an antiferromagnetic order first develops at the edge, 
and later inside the bulk at low temperatures. 
\end{abstract}
\pacs{71.70.Ej, 75.30.Kz, 75.80.+q, 77.80.-e}
\maketitle
The non-trivial topology in condensed matter physics has attracted great interests over the decades 
as highlighted by the celebrated discovery of the quantum Hall (QH) effect~\cite{Prange_Girvin, QHE_thouless82a}.
Haldane theoretically studied the QH effect on the honeycomb lattice even without the Landau levels 
\cite{QHE_haldane88a}, which suggested that the non-trivial topology is more ubiquitous in solids than expected.
A recent breakthrough in this field is the theoretical and experimental discoveries of the quantum 
spin Hall (QSH) effect in time-reversal symmetric insulators~\cite{QSHE_kane05a, QSHE_kane05b, 
QSHE_bernevig06a, QSHE_murakami06a, QSHE_fu06a, QSHE_fu07a, QSHE_qi08c, QSHE_wu06a, QSHE_xu06a, 
QSHE_bernevig06b, QSHE_konig07a}.
Intuitively it can be regarded as two copies of QH systems with up and down spins, but is driven by the spin-orbit interaction (SOI) instead of the external magnetic field.
The $Z_2$ topological number, which distinguishes a topological insulator, i.e. QSH insulator,
from an ordinary band insulator~\cite{QSHE_kane05a}, is closely related to a Kramers doublet protected by the 
time-reversal symmetry, and corresponds to the presence or absence of gapless helical edge modes in the 
semi-infinite system~\cite{QSHE_fu06a}.
The theoretical design of a topological insulator using HgTe/CdTe quantum wells 
\cite{QSHE_bernevig06b} was followed by the 
experimental realization~\cite{QSHE_konig07a}, and now this novel state of matter has 
been firmly established.

However, topological insulators have been limited to semiconductors at low temperature.
This is because we need the large SOI and the fine tuning of the band structure.
Therefore one important development is to realize more robust topological insulators at higher temperature by the larger SOI.
Another interesting development is to study the interplay between the non-trivial topology and the electron correlation~\cite{QSHE_raghu08a}.
Generally the electron correlation is stronger in $d$- and $f$-electrons than in $s$- and $p$-electrons.
When we look at transition metal ions in the periodic table, the electron correlation is the strongest in $3d$ elements and decreases to $4d$ and to $5d$ elements because $d$-orbitals are more and more extended, while the SOI increases as the atomic number.
Thus in $5d$ transition metal oxides, both the SOI and the electron correlation become important with the same order of magnitudes.
In addition, a variety of crystal structures and even tailor-made structures such as superlattice are available in transition metal oxides.
These advantages will be useful to design topological insulators.

In this Letter, we study theoretically the electronic states of a newly synthesized 
compound Na$_2$IrO$_3$~\cite{takagi} in terms of the tight-binding model analysis and
the first-principles calculations as a representative example to 
propose a way to design topological insulators in $5d$ transition metal oxides by 
using the complex transfer integrals and the lattice geometry. This material is predicted to be 
(i) a QSH insulator, (ii) the edge antiferromagnet (AFM), and (iii) the bulk AFM with decreasing temperature.

$5d$-orbitals are rather extended and subject to the large crystalline field.
Under the octahedral crystalline field, $d$-orbitals are split into $e_g~(x^2-y^2,~3z^2-r^2)$- and $t_{2g}~(xy,~yz,~zx)$-orbitals by $10Dq$ of the order of $3{\rm eV}$~\cite{Ir_moon06a}.
The SOI is quenched in $e_g$-orbitals but remains effective in $t_{2g}$-orbitals, which form effectively the triplet with $\ell_{\rm eff} = 1$.
Explicitly, $(\ket{yz} \pm i\ket{zx})/\sqrt{2}$ correspond to $\ket{\ell_{\rm eff}^z = \pm 1}$, while $\ket{xy}$ to $\ket{\ell_{\rm eff}^z = 0}$.
Including the SOI, we obtain the states with the total angular momentum $j_{\rm eff} = 3/2$ and $1/2$.
The wavefunctions with $j_{\rm eff} = 1/2$ read as
\begin{equation}
  \begin{split}
    \ket{+1/2}
    = & (+\ket{xy\uparrow} + \ket{yz\downarrow} + i\ket{zx\downarrow})/\sqrt{3} \\
    \ket{-1/2}
    = & (-\ket{xy\downarrow} + \ket{yz\uparrow} - i\ket{zx\uparrow})/\sqrt{3}.
  \end{split}
  \label{eq:jeff}
\end{equation}
The central idea is that the transfer integrals between these complex orbitals and oxygen orbitals become complex.
For example, consider a $p_z$-orbital.
The transfer integral between $\ket{\pm 1/2}$ and $p_z$ is proportional to $e^{\pm i\theta}$, where $\theta$ is the angle between the $x$ axis and the bond direction.
This complex transfer integral is responsible for topological states in iridates.
Recently, a layered perovskite oxide Sr$_2$IrO$_4$ was studied by angle-resolved photoemission (ARPES), X-ray absorption, optical conductivity, and first-principles calculations~\cite{Ir_kim08a, Ir_kim09a}.
Ir$^{4+}$ has five electrons, one of which is in a narrow band mainly composed by $j_{\rm eff}=1/2$-states 
described above, leading to a Mott insulator with the AFM order.
These experiments confirmed that spin-orbit coupled $j_{\rm eff}=1/2$-states are realized, even though 
Sr$_2$IrO$_4$ itself is topologically trivial.

\begin{figure}
  \centering
  \includegraphics[clip,width=0.45\textwidth]{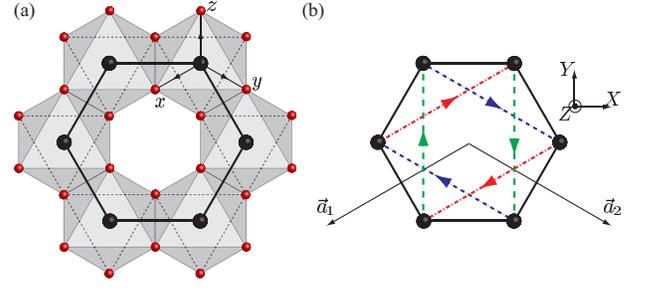}
  \caption{%
  (Color online)
  (a) The honeycomb lattice of Ir atoms in Na$_2$IrO$_3$ viewed from the $c$ axis.
  A large black circle shows an Ir atom surrounded by six O atoms (red small circles).
  (b) The transfer integrals on the honeycomb lattice.
  A black line shows $-t$, while blue dotted, red broken, and green dashed arrows 
indicate $it^{\prime}\sigma_x,~it^{\prime}\sigma_y,~it^{\prime}\sigma_z$, respectively.%
  }
  \label{fig:crystal}
\end{figure}

Now we focus on Na$_2$IrO$_3$ whose layered crystal 
structure contains the honeycomb lattice as shown in Fig.~\ref{fig:crystal} (a) (For the three 
dimensional structure, see Fig. 2(d).)
Each Ir atom is surrounded by an octahedron of six O atoms, which leads to the similar energy level scheme as Sr$_2$IrO$_4$, i.e. one electron in $j_{\rm eff}=1/2$-states.
Therefore we can construct the effective single-band model on the honeycomb lattice.
Since the O $p$-level $\ep$ are around $3{\rm eV}$ lower than the Ir $d$-level $\ed$~\cite{Ir_moon06a}, we can integrate out $p$-orbitals to obtain the following effective Hamiltonian
\begin{equation}
  H_0
  = -t\sum_{\langle ij \rangle}\left[\cre{i}\ann{j} +  \hc\right]
  + \sum_{\langle\langle ij \rangle\rangle}\left[\cre{i}{\hat t}^{\prime}_{ij}\ann{j} + \hc \right],
  \label{eq:latH}
\end{equation}
where $\langle ij \rangle$ and $\langle\langle ij \rangle\rangle$ denote the nearest-neighbor (NN) and next-nearest-neighbor (NNN) pairs, respectively.
The transfer integral $t$ between a NN pair is real and spin-independent as given by
\begin{equation}
  t = \frac{1}{3}\frac{\pdp^2}{\ed - \ep}\frac{\pps + 3\ppp}{\ed - \ep},
  \label{eq:NN}
\end{equation}
where $\pdp$, $\pps$ and $\ppp$ are Slater-Koster parameters between $pd$ and $pp$, respectively~\cite{harrison}.
Note that the contributions of the order of $\pdp^2/(\ed - \ep)$ cancel out in the honeycomb lattice, in sharp contrast to Sr$_2$IrO$_4$ with the square lattice.
The transfer integral between a NNN pair depends on spin, leading to a topological insulator.
The local $x$, $y$, and $z$ axes at an Ir atom are chosen to point in the direction of neighboring O atoms as shown in Fig.~\ref{fig:crystal}.
Therefore $Z = (x + y + z)/\sqrt{3}$ is perpendicular to the honeycomb plane.
With this convention, the transfer integral is a $2 \times 2$ matrix in the spin space, and is written as
\begin{equation}
  {\hat t}^{\prime}_{ij}= it^{\prime}\sigma_a + t^{\prime}_0
  \label{eq:NNNgen},
\end{equation}
where $a = x,~y,~z$ is the direction whose projection onto the honeycomb plane coincides with that of the hopping direction.
The magnitude $t^{\prime}$ is given by
\begin{equation}
  t^{\prime}
  = \frac{1}{6}\frac{\pdp^2}{\ed - \ep}\left[\frac{\pds^2}{(\ed^{\prime} - \ep)^2} + \frac{\pds^2}{(\ed^{\prime} - \ep)(\ed - \ep)}\right]
  \label{eq:NNN}
\end{equation}
with $\ed^{\prime} = \ed + 10Dq$.
Note that the key to these complex transfer integrals is the 
asymmetry between two paths connecting a NNN pair.
If there were an additional Ir atom in the center of the hexagon,
leading to the triangular lattice, the transfer integral $t^{\prime}$ would vanish.
The real transfer integral $t_0^{\prime}$ can be produced by the direct $dd$ hopping and breaks the 
particle-hole symmetry.
However we put $t_0^{\prime} = 0$ for the moment since such term does not change
the topological properties of the Bloch wave functions.

To summarize these results, the transfer integrals are real and spin-independent for a NN pair, 
while complex and spin-dependent for a NNN pair.
We can see that this model is related to the Haldane model for the QH effect~\cite{QHE_haldane88a}, 
and also to the Kane-Mele model for the QSH effect~\cite{QSHE_kane05a, QSHE_kane05b}.
In fact, we can derive the effective Hamiltonian around $K$ and $K^{\prime}$ points of the 
Brillouin zone as
\begin{equation}
  H_0
  = \int d^2r \psi^{\dagger}\left[ 3t^{\prime}\eta_z\tau_z\sigma_Z + \frac{3}{2}t(i\partial_Y\eta_z\tau_x - i\partial_X\tau_y)\right]\psi,
  \label{eq:contH}
\end{equation} 
where $\psi({\vec r})$ is the eight-component spinor field operator, and $k_X$ and $k_Y$ are measured from 
$K$ or $K^{\prime}$ points. $\eta$'s, $\tau$'s, and $\sigma$'s are the Pauli matrices for the valley 
($K$ or $K^{\prime}$), sublattice ($1$ or $2$), and spin ($+$ or $-$) degrees of freedom, respectively.
This effective Hamiltonian is accompanied with the spin Chern number $2$ and belongs to the $Z_2$ non-trivial 
class. However, two spin components are not decoupled away from $K$ or $K^{\prime}$ points, 
and we need to analyze the lattice Hamiltonian Eq.~\eqref{eq:latH} more seriously.
With both time-reversal and inversion symmetries, the Hamiltonian can be written in terms of five 
$4 \times 4$ matrices $\Gamma_a = (\tau_x,~\tau_y,~\tau_z\sigma_x,~\tau_z\sigma_y,~\tau_z\sigma_z)$ 
defined by Fu and Kane~\cite{QSHE_fu07a} as
\begin{equation}
  H_0({\vec k})
  = \sum_{a = 1}^5d_a({\vec k})\Gamma_a
  \label{eq:latHclass}
\end{equation}
with $d_1 = -t(1 + \cos k_1 + \cos k_2)$, $d_2 = -t(-\sin k_1 + \sin k_2)$, $d_3 = 2t^{\prime}\sin k_1$, 
$d_4 = 2t^{\prime}\sin k_2$, and $d_5 = -2t^{\prime}\sin (k_1 + k_2)$.
We can apply the criterion by Fu and Kane~\cite{QSHE_fu07a} and obtain the non-trivial $Z_2$ number 
$(-1)^{\nu} = -1$. In addition, we can directly find a pair of gapless helical edge modes by numerical 
diagonalization for the system with the open boundary condition in one direction.
The crossing point is at $k = \pi$ for the zigzag geometry, while at $k = 0$ for the armchair geometry, 
where $k$ is the wavenumber along the edge.
Anyway such crossing is protected by the Kramers theorem, and can get gapped only if the time-reversal 
symmetry is broken~\cite{QSHE_wu06a}.

\begin{figure}
  \centering
  \includegraphics[clip, width=0.45\textwidth]{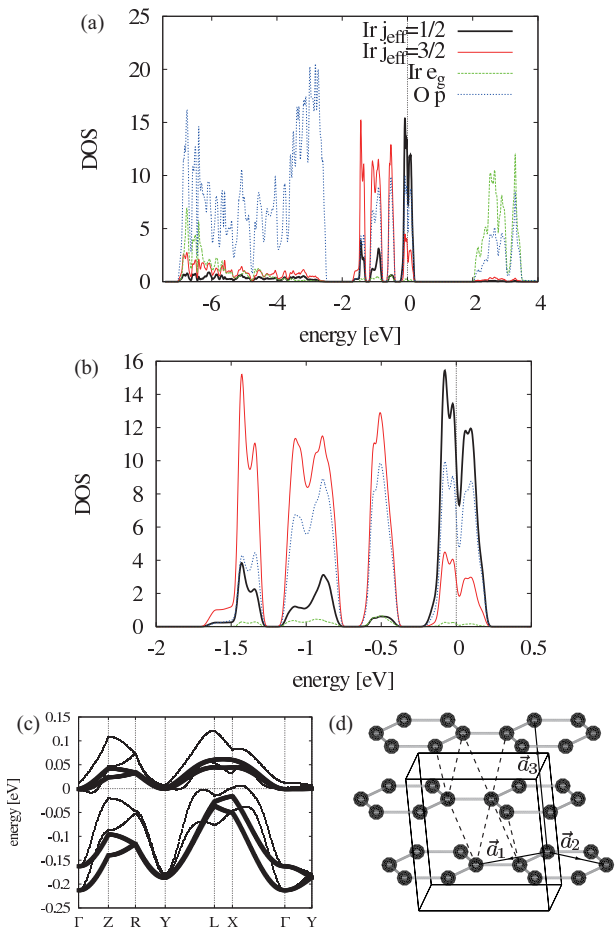}
  \caption{%
  (Color online)
  (a) and (b) The relativistic DOS including the SOI in two different ranges of energy.
  Black thick, red thin, green dashed, and blue dotted lines indicate Ir 
  $j_{\rm eff} = 1/2$-, Ir $j_{\rm eff} = 3/2$-, Ir $e_g$-, and O $p$-bands, respectively.
  The Fermi energy is set to zero.
  (c) The first-principles band structure (thin line) and the extended tight-binding model 
  with typical parameters $t = 310{\rm K}$, $t^{\prime} = 100{\rm K}$, $t_0^{\prime} = -130{\rm K}$, 
  and $t_{\perp} = 60{\rm K}$ (thick line).
  (d) The interlayer coupling $t_{\perp}$ is indicated by black dashed lines, 
  while the other transfer integrals are shown in Fig.~\ref{fig:crystal} (b).
  Due to the monoclinic crystal structure, layers are not stacked in the simple way such as 
  $AB$-stacked graphene.
  }
  \label{fig:first}
\end{figure}

To justify the analysis above based on the 
tight-binding model and to estimate the transfer integrals quantitatively,
we perform the first-principles band structure calculations by using the full-potential linearized 
augumented plane-wave method implemented in WIEN2k~\cite{wien2k}.
Since the accurate crystal parameters of Na$_2$IrO$_3$ are not yet known, we use those of Na$_2$PtO$_3$ 
and then perform atomic position relaxation.
The muffin-tin radii $R_{\rm MT}$ are set to $2.05$, $2.10$, and $1.65{\rm a.u.}$ for Ir, Na, and O atoms, 
respectively. Wavefunctions inside atomic spheres are expanded in spherical harmomics up to 
$\ell_{\rm max} = 10$, while those outside spheres are expanded in plane waves with the cut-off 
$R_{\rm MT}K_{\rm max} = 7.0$.
Figure~\ref{fig:first} (a) shows the relativistic density of states (DOS) including the SOI,
and (b) is its enlargement near the Fermi energy $E_{\rm F} = 0$.
First, the major part of the oxygen $2p$-band is well below $E_{\rm F}$ ranging from $-7$ to $-3{\rm eV}$, but are hybridized with the $5d$-band.
The crystal field splitting between $e_g$- and $t_{2g}$-orbitals is of 
the order of $4{\rm eV}$, which is much larger than that of $3d$-orbitals.
Therefore it is seen that the states near $E_{\rm F}$ is mainly composed 
of the $t_{2g}$-orbitals hybridized with the oxygen $2p$-orbitals, which supports
the assumption of tight-binding analysis described above. 

As shown in Fig.~\ref{fig:first} (b), these $t_{2g}$-orbitals are further 
split by the SOI into $j_{\rm eff}=1/2$- and $j_{\rm eff}=3/2$-states.
From the center of gravity of each of these two partial DOS, 
the splitting is estimated of the order of $1{\rm eV}$, and the 
states near $E_F$ is mostly those of $j_{\rm eff}=1/2$-states. 
There is a finite DOS at $E_{\rm F}$ even though a dip is observed there.
As shown in Fig.~\ref{fig:first} (c), this is due to the overlap of the 
band dispersions of the conduction and valence bands. However, it is 
well known that the local density approximation (LDA) usually underestimates the band 
gap and it is plausible that the gap should open, considering the fact that the  
real material Na$_2$IrO$_3$ is an insulator at room temperature above the 
magnetic phase transition at $T_N \cong 10{\rm K}$.
In Fig.~\ref{fig:first} (c) shown the tight-binding fit (thick line) of the energy dispersions  
obtained by the first-principles band calculation (thin line).
The tight-binding model includes the spin-independent real transfer integral 
$t_0^{\prime}$ between the NNN sites and the interlayer 
transfer $t_{\perp}$ besides the $t$ and $t^{\prime}$ terms in Eq.~\eqref{eq:latH} as shown 
in Fig.~\ref{fig:first} (d).
We set the transfer integrals to $t = 310{\rm K}$, $t^{\prime} = 100{\rm K}$, 
$t_0^{\prime} = -130{\rm K}$, and $t_{\perp} = 60{\rm K}$, and the band structure 
obtained by the extended tight-binding model can roughly reproduce 
the features of the first-principles calculations.
The deviations could be due to the mixing of $e_g$- and $2p$-orbitals.
For the present purpose, i.e., to study the topological nature of the 
Bloch wavefunction, this deviation does not matter. 
Especially, $t_0^{\prime}$ just shifts the energy dispersion and does not
change the Bloch wavefunction itself.  
The helical edge channels for a layer will turn to the 
two Dirac fermions at $k_3 = 0$ and $\pi$ on the surface of 
of the sample due to the non-zero interlayer coupling $t_{\perp}$.
Therefore this system belongs to a ``weak'' topological insulator in the 
classification by Ref.~\cite{QSHE_fu07b}. 
Although the backward scattering between two Dirac fermions could 
induce localization, the ``side'' surface of the sample remains metallic as long as disorder is weak 
because this 2D system belongs to the sympletic class.
Even in the localized case, the finite DOS on the surface including the $c$-axis which can be detected by scanning tunneling spectroscopy (STS), is a distinct feature
from the usual insulator.
Furthermore, the presence of dislocation 
can highlight the non-triviality of a weak topological 
insulator~\cite{QSHE_ran09a}.
To summarize, the tight-binding model analysis above is 
supported by the first-principles calculation provided that the 
band gap is larger than that of LDA estimation.

Up to now we focused on the single-particle band structure, but in the real material the  
electron correlation energy $U$ is of the order of $0.5{\rm eV}$ as 
estimated in Sr$_2$IrO$_4$~\cite{Ir_kim08a}. 
There are two pictures for the effect of the electron correlation
in the present case where two atoms are in the unit cell. One is the weak 
correlation case, where the band insulator can be antiferromagnetically ordered
when $U$ is larger than the critical value.
The other limit is the Mott insulator, where the $U$ localizes an electron at each site
independent of the lattice structure, and only the spin degrees of freedom survive.
It is noted that these two pictures cannot be clearly distinguished and 
cross over as $U$ increases. Therefore it is important to study the two limits
theoretically to understand the real material, which is most probably sitting 
in the intermediate region.

In this Letter, we take the former view that the system is a topological band insulator,
and the magnetic ordering occurs due to the electron correlation at 
low temperature. As discussed above, the nontrivial topology leads to the helical 
edge modes at the boundary of the sample. 
Consequently, it is expected that the magnetic ordering occurs first 
along the boundary of the sample as the temperature is lowered, 
since the gapless helical edge channels are more susceptible to the 
correlation.
Because of the strong SOI, the spin anisotropy is expected.
Within the mean field approximation, we have confirmed that the bulk AFM appears 
below $T_{\rm N}$, and the edge AFM appears just above $T_{\rm N}$, and both of 
these orders are within the $XY$-plane, i.e., the easy-plane anisotropy.
Considering the dispersion along the $c$-axis, the edge AFM moments are coupled 
between layers, leading to the $XY$-like spin model on the surface.
Therefore we expect the Kosterlitz-Thouless transition above the bulk $T_{\rm N}$.
If a single layer of honeycomb lattice can be fabricated, we expect 
the novel one-dimensional correlated electrons along the edge. 
This might be realized by a ``plateau'' on the top surface of the sample.
Since the plateau edge is spatially separated from the edges of underlying layers, 
the interlayer coupling between the edges is reduced and gapless helical edge 
modes are robust.
From the Tomonaga-Luttinger (TL) model analysis of the helical 
edge modes~\cite{QSHE_wu06a,QSHE_xu06a}, it is concluded that the electron-electron 
interaction producing the AFM order and 
the gap in the charge excitation is the 
Umklapp scattering accompanied by the spin flip processes. 
For this interaction to be relevant,
the exponent $K$ should be less than $1/2$, which means the strong repulsive 
interaction~\cite{QSHE_wu06a,QSHE_xu06a}.
For $K > 1/2$, there is no gap opening, and the edge channels remain metallic. 
Therefore the AFM ordering and the metal/insulator transition are always linked 
for the helical edge modes in the weak coupling analysis.
Another interesting aspect is that the domain walls of the edge AFM have 
the fractional charge $\pm e/2$ for each layer~\cite{QSHE_qi08a}.
It is expected that a minigap opening at the edge due to the AFM order
can be detected by STS as temperature is lowered to $\sim T_{\rm N}$, 

In conclusion, we showed that the spin-orbit coupled $d$-orbitals in transition metal 
oxides lead to the topologically non-trivial electronic states.
As an example, the newly synthesized compound Na$_2$IrO$_3$ is found to be a QSH insulator in the 
paramagnetic phase within the tight-binding model as confirmed by first-principles 
calculations.
Three-dimensionality, i.e. the interlayer coupling leads to a layered QSH insulator with 
two Dirac fermions.
At low temperature, the electron correlation drives the system first to the surface AFM phase, 
then to the bulk AFM phase.
Na$_2$IrO$_3$ provides the first experimental stage to study the effect of the electron correlation 
in QSH insulators.

The authors are grateful to H. Takagi and  S. Fujiyama for fruitful discussions.
This work is supported in part by Grant-in-Aids 
(Grant No.~15104006, No.~16076205, No.~17105002, No.~19048015, No.~19048008) 
and NAREGI Nanoscience Project from MEXT.
J.K. was supported by SFB 484 of the Deutsche Forschungsgemeinschaft.
This work is also supported by the NSF under grant numbers DMR-0342832 and the 
US Department of Energy, Office of Basic Energy Sciences under contract DE-AC03-76SF00515.

{\it Note added:} After submission of this paper, we found a paper by Jackeli {\it et al.} 
which suggests that the spin exchange interaction 
in the honeycomb lattice of Ir atoms leads to the Kitaev model~\cite{Ir_jackeli09a}.
Their approach from the strong correlation limit is complementary to our approach from the weak 
correlation limit.

\begin{thebibliography}{10}
\bibitem{Prange_Girvin}
See e.g., The Quantum Hall effect, edited by R.~E. Prange and S.~M. Girvin, (Springer-Verlag, 1987), and references therein.

\bibitem{QHE_thouless82a}
D.~J. Thouless, M. Kohmoto, M.~P. Nightingale, and M. den Nijs, Phys. Rev.
  Lett. {\bf 49},  405  (1982).

\bibitem{QHE_haldane88a}
F.~D.~M. Haldane, Phys. Rev. Lett. {\bf 61},  2015  (1988).

\bibitem{QSHE_kane05a}
C.~L. Kane and E.~J. Mele, Phys. Rev. Lett. {\bf 95},  146802  (2005).

\bibitem{QSHE_kane05b}
C.~L. Kane and E.~J. Mele, Phys. Rev. Lett. {\bf 95},  226801  (2005).

\bibitem{QSHE_bernevig06a}
B.~A. Bernevig and S.-C. Zhang, Phys. Rev. Lett. {\bf 96},  106802  (2006).

\bibitem{QSHE_murakami06a}
S. Murakami, Phys. Rev. Lett. {\bf 97},  236805  (2006).

\bibitem{QSHE_fu06a}
L. Fu and C.~L. Kane, Phys. Rev. B {\bf 74},  195312  (2006).

\bibitem{QSHE_fu07a}
L. Fu and C.~L. Kane, Phys. Rev. B {\bf 76},  045302  (2007).

\bibitem{QSHE_qi08c}
X.-L. Qi, T.~L. Hughes, and S.-C. Zhang, Phys. Rev. B {\bf 78},  195424
  (2008).

\bibitem{QSHE_wu06a}
C. Wu, B.~A. Bernevig, and S.-C. Zhang, Phys. Rev. Lett. {\bf 96},  106401
  (2006).

\bibitem{QSHE_xu06a}
C. Xu and J.~E. Moore, Phys. Rev. B {\bf 73},  045322  (2006).

\bibitem{QSHE_bernevig06b}
B.~A. Bernevig, T.~L. Hughes, and S.-C. Zhang, Science {\bf 314},  1757
  (2006).

\bibitem{QSHE_konig07a}
M. K{\" o}nig {\it et~al.}, Science {\bf 318},  766  (2007).

\bibitem{QSHE_raghu08a}
S. Raghu, X.-L. Qi, C. Honerkamp, and S.-C. Zhang, Phys. Rev. Lett. {\bf 100},
  156401  (2008).

\bibitem{takagi}
H. Takagi, private communication.

\bibitem{Ir_moon06a}
S.~J. Moon {\it et~al.}, Phys. Rev. B {\bf 74},  113104  (2006).

\bibitem{Ir_kim08a}
B.~J. Kim {\it et~al.}, Phys. Rev. Lett. {\bf 101},  076402  (2008).

\bibitem{Ir_kim09a}
B.~J. Kim {\it et~al.}, Science {\bf 323},  1329  (2009).

\bibitem{harrison}
W.~A. Harrison, Elementary Electronic Structure (World Scientific, Singapore, 1999).

\bibitem{wien2k}
P. Blaha {\it et~al.}, {\em {An Augmented Plane Wave $+$ Local Orbitals Program
  for Calculating Crystal Properties}} (Karlheinz Schwarz, Techn. Universit{\"
  a}t Wien, Austria, 2001).

\bibitem{QSHE_fu07b}
L. Fu, C.~L. Kane, and E.~J. Mele, Phys. Rev. Lett. {\bf 98},  106803  (2007).

\bibitem{QSHE_ran09a}
Y. Ran, Y. Zhang, and A. Vishwanath, Nat. Phys. {\bf 5},  298  (2009).

\bibitem{QSHE_qi08a}
X.-L. Qi, T.~L. Hughes, and S.-C. Zhang, Nat. Phys. {\bf 4},  273  (2008).

\bibitem{Ir_jackeli09a}
G. Jackeli and G. Khaliullin, Phys. Rev. Lett. {\bf 102},  017205  (2009).

\end{thebibliography}

\end{document}